\documentclass{cimento}

\usepackage{graphicx}  % got figures? uncomment this
\usepackage{graphicx,wrapfig,lipsum}
\usepackage{verbatim}
\title{Proton-proton and proton-cluster femtoscopy at the HADES  experiment}

\author{M. Stefaniak for HADES Collaboration\from{ins:x}\from{ins:y}}

\instlist{
\inst{ins:x} Department of Physics, The Ohio State University, Columbus,  USA; 
\inst{ins:y} GSI Helmholtz Centre for Heavy Ion Research, Darmstadt, Germany;}

\begin{document}

\maketitle

\begin{abstract}
This work explores femtoscopic correlations in proton-proton and proton-cluster systems at the HADES experiment, GSI. Through high-precision correlation functions, it reveals the impact of strong interactions between protons and light nuclei (deuteron, triton, Helium-3). For proton-proton interactions, measurements in two centralities and five $k_T$ bins were analyzed. Extracted radii were compared with different Equations of State in UrQMD, providing valuable insights into collision dynamics. This research enhances our understanding of particle interactions and contributes to refining theoretical models in nuclear and particle physics.

\end{abstract}

\section{Introduction}

The exploration of the Quantum Chromodynamics phase diagram stands as a central and ambitious goal for numerous nuclear facilities. A recent surge in interest has been directed towards a region characterized by finite net-baryon densities and relatively low temperatures. This focus is driven by the belief that the study of heavy-ion collisions within this specific domain can offer valuable insights, not only for nuclear physics but also as a reference for astrophysical investigations into Neutron Star Mergers (NSM). Such studies are pivotal for unraveling the intricacies of the Equation of State (EoS) governing nuclear matter.

One of the key challenges in this pursuit lies in comprehending the Strong Interactions (SI) between hadrons and the excited or unbound states of nuclear matter. These interactions remain inadequately described, posing a significant gap in our understanding of the EoS. To address this, the HADES experiment emerges as a crucial player, employing a specialized tool known as Femtoscopic Correlations. This approach allows for a detailed examination of the complex interaction of particles through precise correlation functions.

\section{HADES experiment}

The \textbf{H}igh \textbf{A}cceptance \textbf{D}i\textbf{E}lectron \textbf{S}pectrometer (HADES) stands as a fixed-target experiment situated at the SIS-18 accelerator within the GSI Helmholtz Centre for Heavy Ion Research in Germany. This state-of-the-art experiment boasts an almost complete azimuthal acceptance and covers polar angles ranging from $18^0$ to $85^0$. This broad acceptance enables HADES to comprehensively capture particle emissions from heavy-ion collisions, offering a unique perspective on the intricate dynamics within these interactions.

The created systems in these collisions can be characterized by temperatures ($T$ = 60-80 MeV) and net baryon densities ($\rho$ $<$ 2-3 $\rho_0$), closely resembling the thermal and density parameters observed in NSM scenarios \cite{HADES:2019auv}. This selection of collision parameters enhances the applicability of our findings to astrophysical studies, contributing valuable insights to the determination of the EoS of nuclear matter.

\section{Femtoscopic correlations}

Femtoscopic correlations (FC) serve as a powerful tool for probing the geometric and dynamic properties of the particle-emitting source \cite{FemtoLisa}. By delving into these correlations, we gain valuable insights into potential excited states of particles, extracting crucial information about the SI between the involved particles. The theoretical framework for femtoscopic correlations is often encapsulated by the Koonin-Pratt formula \cite{FemtoLisa}:

\begin{equation}
    C(k^{\star}) = \int S(r^{\star})|\Psi(k^{\star},r^{\star})|^2 d^3r^{\star}
\end{equation}

Here, \(S(r^{\star})\) represents the source function, describing the distribution of the relative positions (\(r^{\star}\)) of the two particles, while \(\Psi(k^{\star},r^{\star})\) is the two-particle wave function. Derived from the Schrödinger equation for a given potential characterizing particle interactions, the wave function is a key element encoding the effect of two-particle interactions on femtoscopic correlations. Finally, \(k^{\star}\) denotes the relative momentum of the particles in the pair rest frame.

The femtoscopic correlation function's shape is influenced by the convolution of three primary effects:

- \textbf{Coulomb interaction:} This factor depends on the electric charge of the particles and contributes to shaping the correlation function: attractive unlike-sign particles, repulsive same-sign particles.
  
- \textbf{Quantum statistics:} Governed by the Pauli exclusion law, quantum statistics introduce a repulsion for fermions, influencing negatively overall form of the correlation function.
  
- \textbf{Strong Interactions:} The femtoscopic correlation function is profoundly influenced by the potentials between correlated particles, encompassing both attractive and repulsive forces. The presence of resonances further contributes to the varied forms of the correlation function, adding complexity to the analysis.

%\begin{figure}
%\centering
%\includegraphics[scale=0.4]{corr_func_pp.eps}
%\caption{Proton-Proton Correlation Function. The correlation function, generated using CorAl software \cite{coral}, illustrates the interplay of Coulomb interaction, quantum statistics (green), and SI (red).}
%\label{fig:corr_func_pp}
%\end{figure}

%For proton-proton correlations, we provide a graphical representation of the correlation function in Figure \ref{fig:corr_func_pp}. The green curve illustrates the influence of Coulomb interaction and quantum statistics, resulting in a repulsive force. In contrast, the red curve incorporates the additional effect of Strong Interactions (SI), showcasing a strong positive peak indicative of attractive correlations.

Experimentally, the femtoscopic correlation function is based on the relative momenta of particle pairs and can be expressed as \(C(k^{\star}) = A(k^{\star}) / B(k^{\star})\). Here, \(A(k^{\star})\) and \(B(k^{\star})\) represent the distributions of \(k^{\star}\) for all possible pairs within one event and for mixed events, respectively. The latter serves as a representation of uncorrelated particles, and the division helps suppress non-femtoscopic correlations arising from factors such as flow, non-uniform detector efficiency, and other extraneous influences \cite{FemtoLisa}.

\section{Results}
In this section, we present the results derived from an analysis of data obtained by the HADES experiment for Ag+Ag collisions at a center-of-mass energy of $\sqrt{s_{NN}}$ = 2.55 GeV. 

To address potential experimental effects, we implemented the following measures. The challenges posed by two-track tracking detector effects, particularly merging, were effectively mitigated by the application of a $\Delta \theta > 0.04 $ and $\Delta\phi > 0.25$ cut. 

Flow effects, a common phenomenon in heavy-ion collisions, were systematically suppressed by adopting mixing events with similar event planes. This is achieved by requiring the respective event plane angles to the in the same intervals of 30 degree width.

Additionally,an energy loss correction was applied to the reconstucted tracks to account for variations in particle energies due to interactions with the medium.

\subsection{Proton-light nuclei correlation functions}

\begin{figure}
\centering
\includegraphics[scale=0.4]{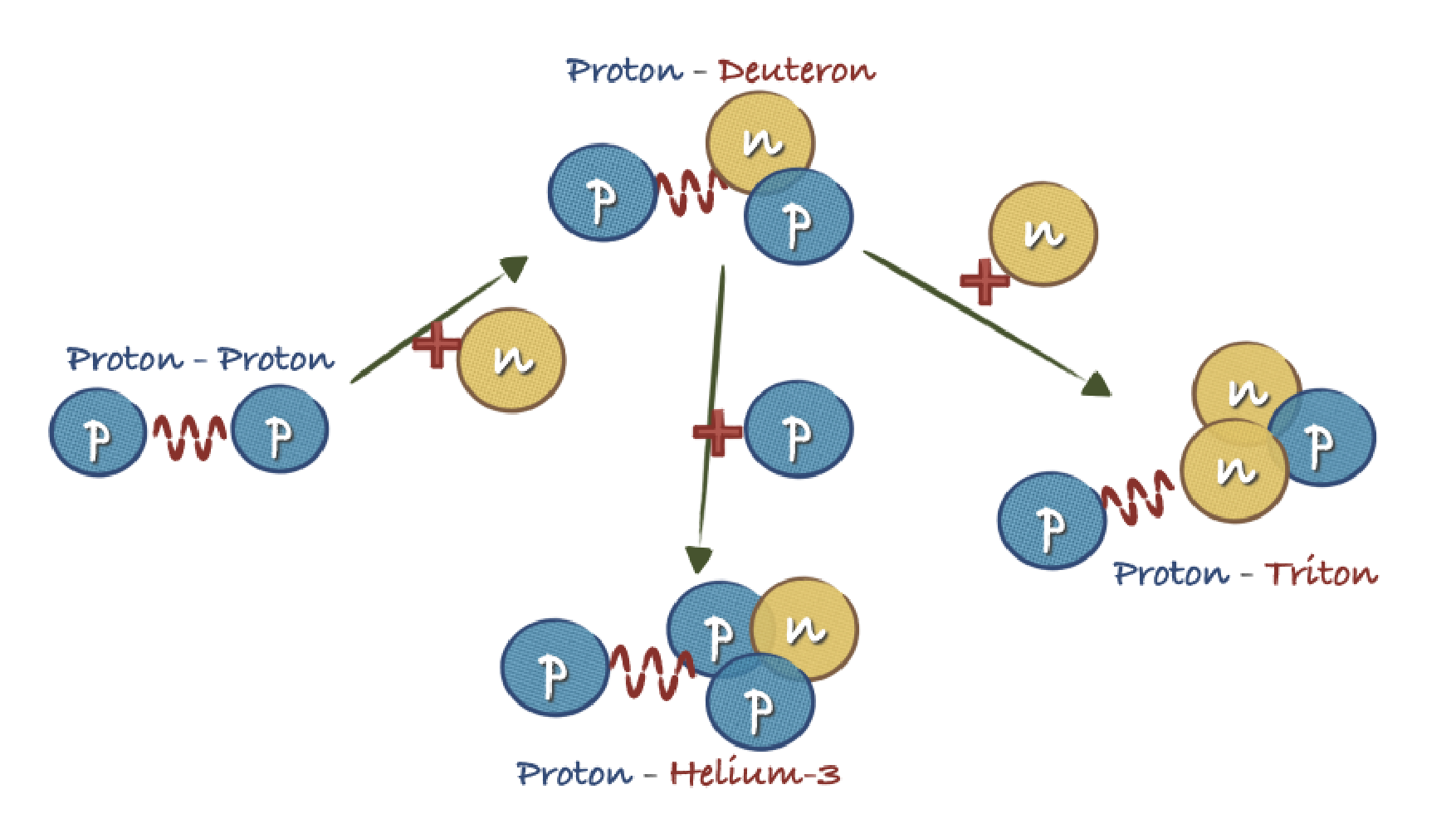}
\caption{Sequence of analyzed proton-light nuclei systems.}
\label{fig:corrPairs}
\end{figure}

\begin{figure}
\centering
\includegraphics[scale=0.5]{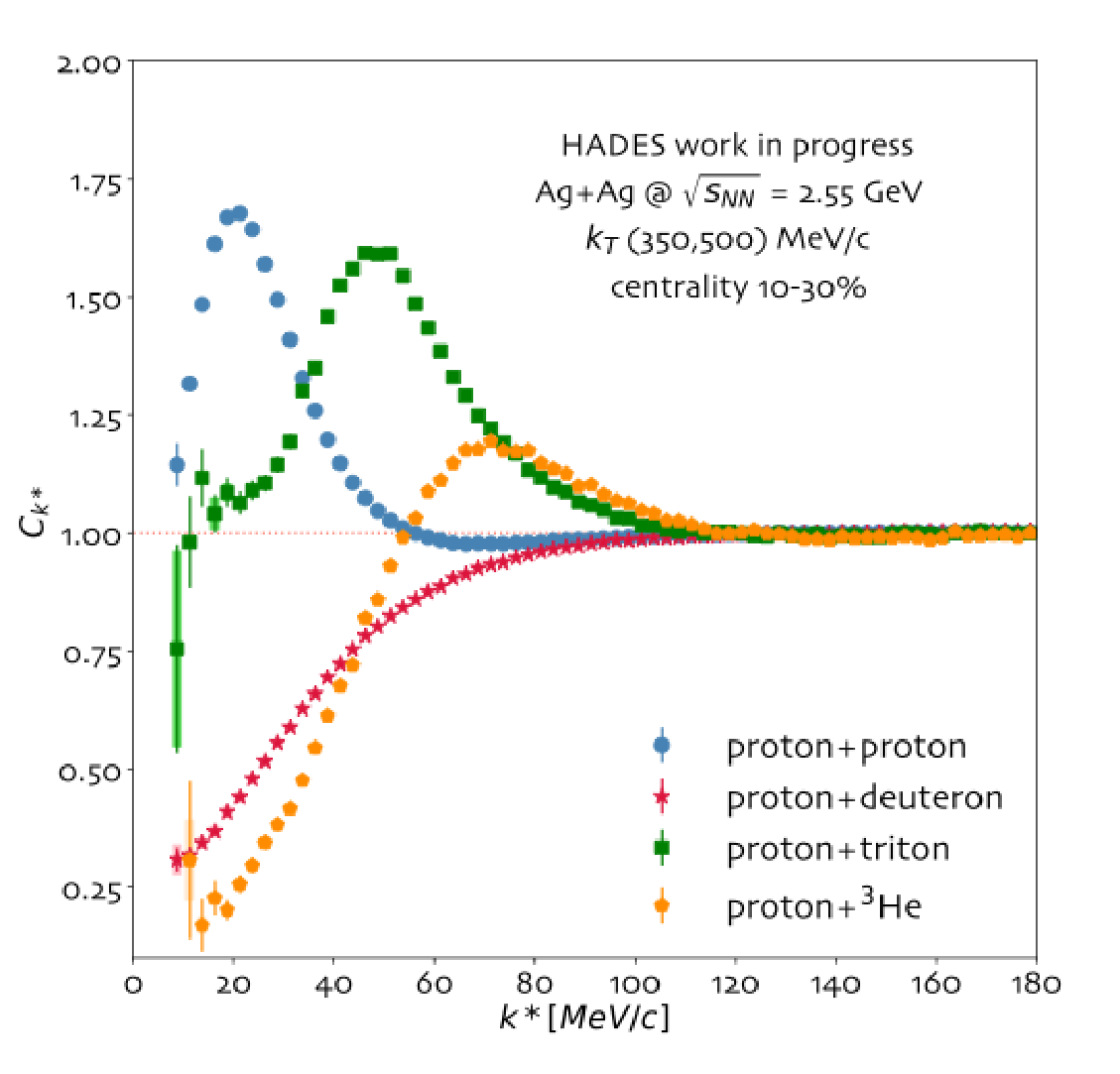}
\caption{Measured correlation functions of p-p (blue), p-d (red), p-t(green), p-$^3$He (yellow) in Ag+Ag collisions at $\sqrt{s_{NN}}$ = 2.55 GeV.}
\label{fig:p_ln_pairs_result}
\end{figure}

We conducted a series of measurements involving the correlation functions of proton-light nuclei systems. The sequence began with proton-proton (p-p) correlations, followed by the addition of a neutron to form proton-deuteron (p-d), and subsequently, the inclusion of an additional neutron to create proton-triton (p-t), respectively of a  proton to form proton-helium-3 (p-$^3$He) systems (see Figure \ref{fig:corrPairs}). These measurements were carried out within the $10\%-30\%$ centrality range in the transverse momentum ($k_T$) bin of 350-500 MeV/c  and are presented in Fig. \ref{fig:p_ln_pairs_result}.

The proton-proton correlation, represented in blue, exhibits a positive correlation attributed to SI. Additionally, it features a negative correlation induced by the repulsive Coulomb interaction and Quantum Statistics, aligning with theoretical expectations.

In the case of proton-deuteron correlations, depicted in red, the introduction of a neutron to the correlated system results in repulsive interactions between p-d pairs. Our correlation measurements are in good agreement with previous FOPI results but also provide superior precision when compared to other published results \cite{fopi,kemi}. 

Moving to proton-triton correlation, depicted in green, the inclusion of an additional neutron in the correlated system (forming p-t) yields a strong positive correlation. This correlation displays a sharp peak, possibly linked to light nuclei decays, with greater prominence compared to FOPI results.

Finally, for p-$^3$He correlation, shown in yellow, the addition of a proton to the correlated system (p-$^3$He) reveals a strong positive correlation. As there is no comparison  to FOPI shown in Fig. \ref{fig:p_ln_pairs_result} this statement remains unclear and I would suggest to remove it here.

\begin{comment}
\begin{enumerate}
    \item p-p (in blue):
    \begin{itemize}
        \item Exhibits a positive correlation attributed to Strong Interactions.
        \item Features a negative correlation induced by Coulomb and Quantum Statistics, aligning with theoretical expectations.
    \end{itemize}
    \item p-d (in red):
    \begin{itemize}
        \item Introduction of a neutron to the correlated system results in repulsive interactions between proton-deuteron pairs.
        \item Demonstrates superior measurement precision compared to other published results \cite{fopi,kemi}, resembling FOPI findings with enhanced accuracy. 
    \end{itemize}
    \item p-t (in green):
    \begin{itemize}
        \item Inclusion of an additional neutron in the correlated system (forming p-t) yields a strong positive correlation.
        \item Displays a sharp peak, possibly linked to light nuclei decay, with greater prominence compared to FOPI results.
    \end{itemize}
    \item p-3He (in yellow):
    \begin{itemize}
        \item Addition of a proton to the correlated system (p-$^3$He) reveals a strong positive correlation.
        \item A subtle but discernible positive enhancement sets it apart from FOPI's data.
        \item Our measurements underscore the clear impact of Strong Interactions, offering superior precision in comparison to FOPI results.
    \end{itemize}
\end{enumerate}

\end{comment}

\begin{figure}
\centering
\includegraphics[scale=0.5]{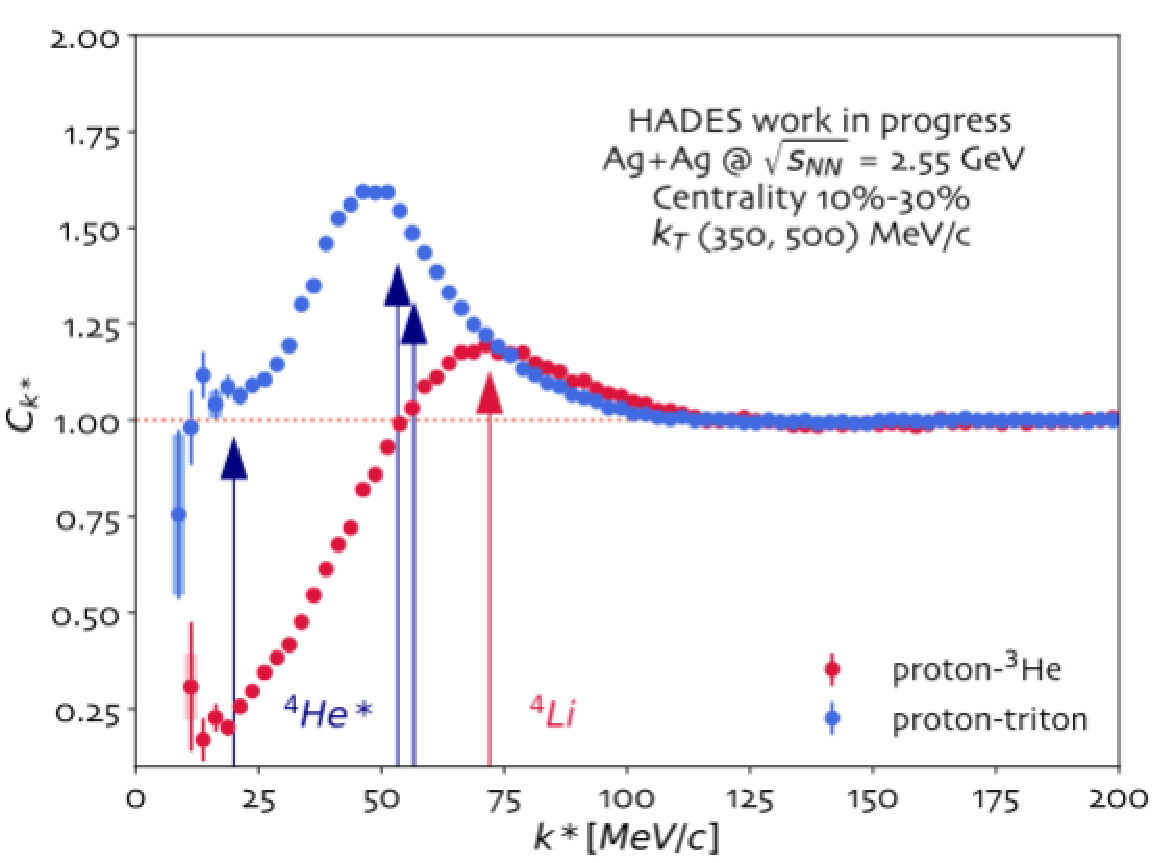}
\caption{Proton-triton (blue) and proton-$^3$He (red) correlation functions. Arrows correspond to positions of the possible decays of nuclear states.}
\label{fig:decays}
\end{figure}

In Figure \ref{fig:decays}, correlation functions for p-t and p-$^3$He pairs are compared. Despite triton and $^3$He having similar masses and consisting of three nucleons, their correlation functions exhibit significant differences. The observed variations in correlation function values are attributed to the decays of excited nuclear states. The specific decays are summarized as follows:
\\
\textbf{$^4 He^{\star} \rightarrow p + t $:}
\begin{itemize}
    \item E = 20.21 MeV, $J_\pi$ = $0_+$, $\Gamma$ = 0.5 MeV, $\Gamma_p/\Gamma$ = 1, $k_1^{\star}$ = 20 MeV/c
    \item E = 21.01 MeV, $J_\pi$ = $0_+$, $\Gamma$ = 0.84 MeV, $\Gamma_p/\Gamma$ = 0.76, $k_2^{\star}$ =53.3 MeV/c
    \item E = 21.84 MeV, $J_\pi$ = 2, $\Gamma$ =2.01 MeV, $\Gamma_p/\Gamma$ = 0.63, $k_3^{\star}$ =  56.6 MeV/c
\end{itemize}
\textbf{$^4 Li \rightarrow p + ^3 He$:}
\begin{itemize}
    \item $J_\pi$ = 2-, $\Gamma$ = 6.0 MeV,$\Gamma_p/\Gamma$ = 1, $k_1^{\star}$  72 MeV/c
\end{itemize}

While the enhancements in the correlation functions are visibly apparent, estimating the number of decaying light nuclei is non-trivial due to the convolution of repulsive Coulomb interaction and attractive SI shaping the correlation function's profile.

\subsection{Proton-proton correlation functions}

\begin{figure}
\centering
\includegraphics[scale=0.4]{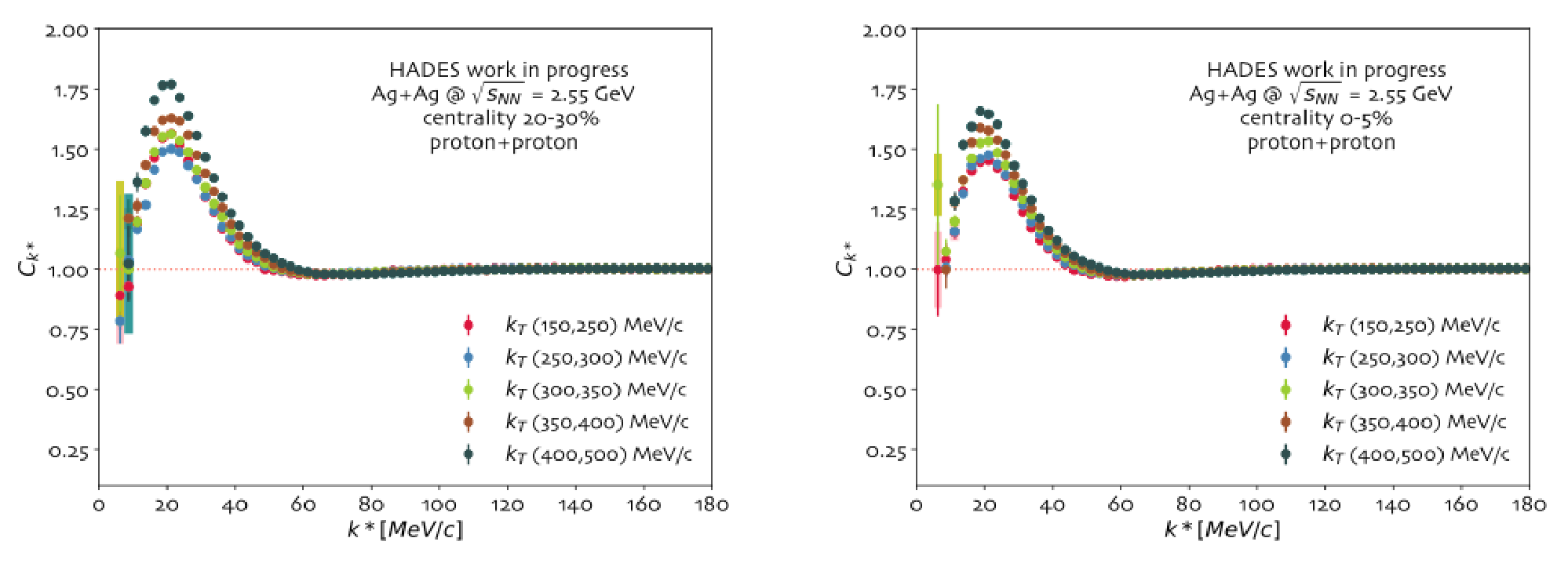}
\caption{Proton-proton correlation functions at $20\%-30\%$ (left panel) and $0-5\%$ (right panel) centralities.}
\label{fig:pp_corr}
\end{figure}

We present experimental results obtained for two centralities ($20\%-30\%$ and $0-5\%$) and across five $k_T$ bins, as illustrated in Figure \ref{fig:pp_corr}. The high-quality statistics in our dataset enables thorough investigations of the dependence on $k_T$.
As $k_T$ increases, the correlation strength amplifies, implying a reduction of the size (R) of the particle-emitting source. 
At lower $k_T$ values, the correlation signals are notably influenced by the resolution of momentum measurements and a less effective suppression of track merging, causing an increase of systematic uncertainties.

\subsubsection{\textit{Femtoscopy vs EoS}}

\begin{figure}
\centering
\includegraphics[scale=0.45]{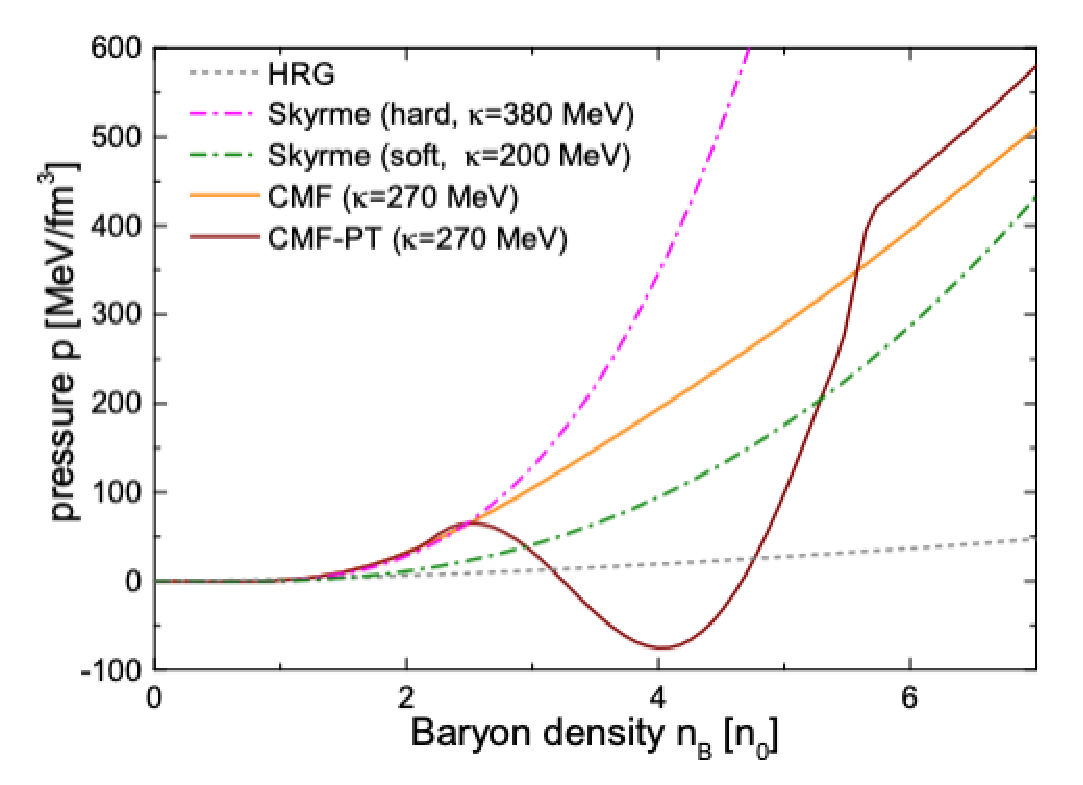}
\caption{Pressure dependence on net baryon density of five EoS implemented in UrQMD. \textit{HRG} correspond to cascade mode, \textit{Soft} and \textit{Hard} EoS are based on Skyrme potential, while \textit{CMF} is chiral magnetic field, and \textit{CMF-PT} includes the phase transition \cite{jan3}.}
\label{fig:potentials}
\end{figure}

In these simulations, depicted in Figure \ref{fig:potentials}, different EoS were implemented using the hadronic transport code UrQMD \cite{urqmd1, urqmd2}. For further insights into the specific characteristics of the employed EoS, readers are directed to the details provided in \cite{jan1, jan2}. The EoS and simulations were prepared by Jan Steinheimer \cite{jan3}.

The primary aim of these simulation studies is not to definitively establish the ultimate EoS. Instead, the focus is on examining how different EoS relate to femtoscopic measurements of p-p pairs. To achieve this, extreme scenarios were considered; for instance, a soft EoS is characterized by a remarkably low stiffness ($\kappa$ = 200 MeV), while an EoS incorporating a phase transition (CMF-PT) is accentuated to a high degree. 

It's worth noting that while analogous studies have been conducted for pion-pion pairs \cite{jan2}, this investigation extends the scope to the realm of p-p femtoscopy, offering novel insights into the interplay between EoS and particle correlations.

The theoretical correlation functions were derived using the CorAl software \cite{coral}, a development of Scott Pratt. This software incorporates the Reid potential from the referenced work \cite{reid} in the calculation of the pair wave function. Additionally, the source function is constructed by utilizing the output from UrQMD simulations. This dual integration of theoretical tools ensures a comprehensive and accurate representation of the correlation functions, taking into account both potential interactions and simulated source characteristics.

\begin{figure}
\centering
\includegraphics[scale=0.38]{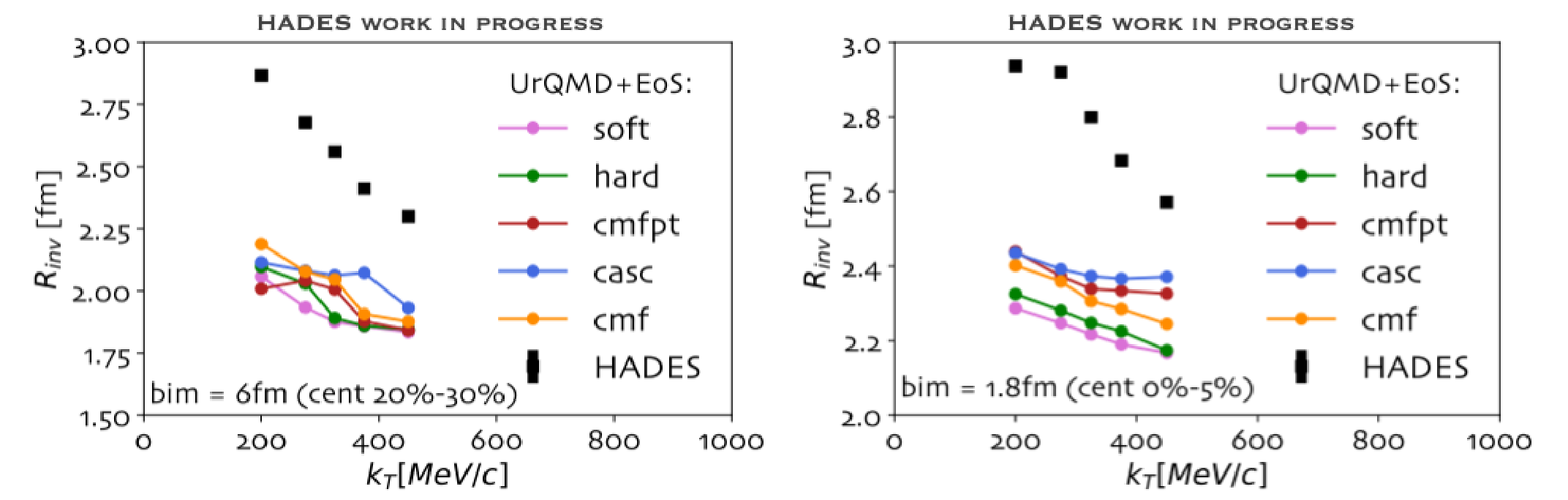}
\caption{Particle emitting source dependence on $k_T$ for experimental  (no systematic uncertainties included) and simulated data. Left panel correspond to midcentral collisions ($20\%-30\%$) and right one to central ($0-5\%$).}
\label{fig:radius}
\end{figure}

The source radius is extracted by fitting correlation functions using CorAl, assuming a Gaussian source. In Figure \ref{fig:radius}, the comparison between HADES experimental data (black points) and UrQMD simulations with different EoS for two centralities reveals consistently larger source radii in the experimental data. The discrepancies between experimental and simulated source radii can be attributed to the exclusion of light nuclei production in the simulations, challenges in accurately estimating proton "feed-down" contributions (e.g., from $^5$Li), and the assumption of a Gaussian source in the fitting procedure.

In midcentral collisions, where differences between EoS are relatively small. They start to become more pronounced as we transition to central collisions with higher net baryon densities. Significantly, the comparison reveals notable differences between the various EoS. The Skyrme EoS stands out as the furthest from the experimental data.

\section{Conclusions}

In conclusion, our study on proton-light nuclei interactions has yielded precise measurements, revealing the intricate effects of SI between protons and light nuclei. The investigation further uncovered and the influence of decays of excited states, specifically $^4$He$\star$ and $^4$Li.
Our investigation into proton-proton interactions revealed a robust and pronounced dependence of correlation functions on $k_T$. This observation underscores the significance of the $k_T$ parameter in shaping the dynamics of p-p collisions, providing valuable insights into the complexities of these interactions.
The comparative analysis of different EoS within simulation revealed shared features. Intriguingly, while there were nearly no discernible differences between EoS when considering variations in centrality within the same collision energy, the transition to higher baryon densities in more central events accentuated more pronounced distinctions. These findings underscore the sensitivity of correlation functions to changes in EoS, providing valuable insights for refining EoS models in the context of high-energy proton-proton collisions.

\acknowledgments
I express sincere gratitude to Scott Pratt for invaluable support and to Jan Steinheimer for providing simulated data with EoS along with engaging in constructive discussions.
This work was supported by Humboldt Foundation grant for postdoctoral fellows and U.S. Department of Energy grant DE-SC0020651.

\end{document}